# Exploring Mechanisms of Hydration and Carbonation of MgO and Mg(OH)$_2$ in Reactive Magnesium Oxide-based Cements

Mina Ghane Gardeh,[a] Andrey A. Kistanov,[*,b] Hoang Nguyen,[a] Hegoi Manzano,[c] Wei Cao,[b] and Paivo Kinnunen[a]

[a]Fibre and Particle Engineering Research Unit, University of Oulu, Pentti Kaiteran katu 1, 90014 Oulu, Finland.
[b]Nano and Molecular Systems Research Unit, University of Oulu, Pentti Kaiteran katu 1, 90014 Oulu, Finland.
[c]Departament of Condensed Matter Physics, University of the Basque Country (UPV/EHU), Barrio Sarriena, s/n, 48940 Leioa, Spain.

**\*Corresponding author:** andrey.kistanov@oulu.fi

**Abstract**
Reactive magnesium oxide (MgO)–based cement (RMC) can play a key role in carbon capture processes. However, knowledge on the driving forces that control the degree of carbonation and hydration and rate of reactions in this system remains limited. In this work, density functional theory–based simulations are used to investigate the physical nature of the reactions taking place during the fabrication of RMCs under ambient conditions. Parametric indicators such as adsorption energies, charge transfer, electron localization function, adsorption/dissociation energy barriers and the mechanisms of interaction of H$_2$O and CO$_2$ molecules with MgO and brucite (Mg(OH)$_2$) clusters are considered. The following hydration and carbonation interactions relevant to RMCs are evaluated: *i*) carbonation of MgO, *ii*) hydration of MgO, carbonation of hydrated MgO, *iii*) carbonation of Mg(OH)$_2$, *iv*) hydration of Mg(OH)$_2$ and *v*) hydration of carbonated Mg(OH)$_2$. A comparison of the energy barriers and reaction pathways of these mechanisms shows that the carbonation of MgO is hindered by presence of H$_2$O molecules, while the carbonation of Mg(OH)$_2$ is hindered by the formation of initial carbonate and hydrate layers as well as presence of excessed H$_2$O molecules. To compare these finding to bulk mineral surfaces, the interactions of the CO$_2$ and H$_2$O molecules with the MgO(001) and Mg(OH)$_2$ (001) surfaces are studied. Therefore, this work presents deep insights into the physical nature of the reactions and the mechanisms involved in hydrated magnesium carbonates production that can be beneficial for its development.

**Introduction**

Increasing carbon dioxide ($CO_2$) emissions are currently one of the most serious environmental challenges.[1] Cement manufacturing, and specifically the manufacture of ordinary Portland cement (OPC), is the source of ~5%–7% of global greenhouse gas emissions.[2] Limestone ($CaCO_3$), the conventional feedstock for OPC manufacturing, is excavated, crushed and sintered with other materials in a cement kiln at temperatures reaching ~1450°C to produce clinker. During the calcination of $CaCO_3$, $CO_2$ is directly emitted (i.e. $CaCO_3 \rightarrow CaO + CO_2$), causing ~50%–60% of the total emissions from OPC production.[3] From the standpoint of sustainable development, the cement industry is seeking alternatives to reduce $CO_2$ emissions while maintaining the same performance.[4]

Among the proposed alternative binders, Mg-based cements have attracted attention for their promise as partial replacements for OPC.[5] When magnesium oxide (MgO) is derived from Mg silicates (e.g. olivine and serpentine), less environmental and economic impact is generated.[6] The net $CO_2$ emissions from the carbonation of these binders may be ~73% lower than OPC,[7] and, therefore, may potentially lead to the formation of carbon-negative cements. Moreover, the lower production temperature of reactive MgO–based cement (RMC) compared to that of OPC (i.e. 700–1000°C vs. 1450°C), and its potential to gain strength through its reaction with $CO_2$, has attracted special attention.[7]

Considering the need for the rapid development of carbon capture and utilisation technology,[8] the main advantage of RMCs produced from Mg-Si minerals in concrete formulations is their ability to absorb and permanently store $CO_2$ in the form of stable carbonates during the carbonation process, when MgO is sourced from low-$CO_2$ feedstocks.[9] In such processes, MgO reacts with water ($H_2O$) to form brucite ($Mg(OH)_2$), which generally has a weak and porous structure.[5,10] However, hydrated MgO has a strong ability to absorb $CO_2$ and produce carbonated products at a strength useful for construction purposes.[11] In other words, the dissolution of MgO through hydration results in the formation of $Mg(OH)_2$, which is then carbonated according to the following reaction and produces a range of hydrated magnesium carbonates (HMCs): $Mg(OH)_2 + CO_2 + 2H_2O \rightarrow MgCO_3 \cdot 3H_2O$. Nesquehonite ($MgCO_3 \cdot 3H_2O$) is the most commonly obtained HMC, yet other phases such as hydromagnesite ($4MgCO_3 \cdot Mg(OH)_2 \cdot 4H_2O$), dypingite ($4MgCO_3 \cdot Mg(OH)_2 \cdot 5H_2O$) and artinite ($MgCO_3 \cdot Mg(OH)_2 \cdot 3H_2O$) can also be present.[12,13]

Recent experimental studies have examined the formation of HMCs through the hydration and carbonation of RMC. In particular, improvement of the hydration and mechanical performance of carbonated MgO-based systems has been observed with the introduction of various hydration agents at different concentrations.[14] In this way, the simultaneous use of magnesium acetate at 0.05 M and carbonate seeds (up to 1% of cement content) improved mechanical performance of carbonated RMC concrete mixes.[15] However, investigation of the physical nature of mechanisms involved in the reactions of HMC production is still immature. One of the reasons for this is the limitation of available experimental methods for the determination of such processes occurring at the nanoscale in bulk materials.

Theoretical approaches with predictive capabilities, such as those based on the density functional theory (DFT), show a high capability for determining the most stable atomic structures and exploring the physical and chemical properties of these finite systems.[16,17,18,19] Computational approaches have been successfully utilised to investigate in depth the mechanisms related to the formation of HMCs. For instance, the structure, formation energy and electronic properties of four commonly exposed surfaces of nesquehonite crystal have been studied using DFT-based calculations.[16] In another computational work the activity and selectivity of MgO surfaces for $CO_2$ conversion have been studied.[20] In particular, the adsorption and dissociation of $CO_2$, as well as its subsequent hydrogenation to HOCO and HCOO, on various MgO surfaces, have been investigated. It has been shown that the direct dissociation of $CO_2$ on MgO is thermodynamically unfavourable because of high reaction energy, while hydrogenation of $CO_2$ to HCOO by hydride H is more feasible on MgO. DFT simulations have also been utilised to compare the adsorption and activation reaction mechanisms of $CO_2$ and $H_2$ molecules on hydrogen-assisted MgO(110), pure Ni(111) and Ni/MgO interfaces.[21] Computational methods have also been applied for a deeper exploration of the effects of various promoters and dopants upon $CO_2$ adsorption on the MgO−CaO(100) surface.[22] Theoretically supported experimental infrared-based studies have been performed to identify the structure of the $CO_2$ species adsorbed on the various MgO surface.[23] It has been shown that the active site towards $CO_2$, which is a Lewis acid, differs from that for the deprotonating adsorption of Brønsted acids. Another experimentally supported computational study provided a comprehensive study on the $CO_2$ adsorption on the MgO and $Mg(OH)_2$ surfaces.[24] It has been found that chemisorption of $CO_2$ on the MgO surface is facilitated by the presence of $H_2O$.

Since the reaction degrees of MgO and $Mg(OH)_2$ are relatively low (ca. 50%), they reduce the effectiveness of $CO_2$ utilization to form a cementitious binder.[25] Furthermore, because the transformation of HMCs shows mixed diffusion and reaction-limited control, and it proceeds through the production of metastable intermediates, the specifics of nesquehonite conversion to other HMCs remains unclear. The conversion of these metastable intermediates also raises concerns about the durability of cement.[26] Therefore, insights into the potential reactions in the $MgO/H_2O/CO_2$ system, and an understanding of the nature of kinetic hindrance in MgO and $Mg(OH)_2$ carbonation and hydration at the atomic level, are of immediate interest.

In this work, the physical nature of the mechanisms for HMC production on MgO and $Mg(OH)_2$ nanoclusters is considered using DFT calculations. Clusters are collections of atoms that act as a link between gases and bulk phase materials (liquids and solids). They are considerably large to be considered as molecules while considerably small to be classified as liquids or solids, and almost all of the atoms in a cluster are on or near its surface, making them a good choice for considering surface reactions.[27] In addition, robust reactions at oxide surfaces, such as the exchange rates of $H_2O$ molecules on the surface can be reliably predicted using molecular-simulation methods.[28]

Here, the interaction of these nanoclusters of potentially promising RMC raw materials with ambient molecules ($H_2O$ and $CO_2$) is considered. The mechanism of the following reactions is investigated: carbonation of MgO, hydration of MgO, carbonation of hydrated MgO, carbonation of $Mg(OH)_2$, hydration of $Mg(OH)_2$ and hydration of carbonated $Mg(OH)_2$. Notably, even though through-solution dissolution–precipitation reactions are often the dominating reactions in HMC synthesis, surface carbonation can become important to the overall carbonation kinetics by hindering further reactions, including dissolution. Understanding the mechanisms of these reactions is accomplished by calculating adsorption energy, charge transfer, electron localization function and adsorption/dissociation energy barriers of $H_2O$ and $CO_2$ upon reactions with the MgO and $Mg(OH)_2$ clusters. To gain further insights into the difference between MgO and $Mg(OH)_2$ clusters and bulks, the interactions between the surfaces of bulk MgO and $Mg(OH)_2$ with $H_2O$ and $CO_2$ molecules are also investigated. The results also shed light on the underlying reason for the hindrance of carbonation of MgO and $Mg(OH)_2$ that has been previously observed experimentally. Therefore, the results of this work reveal the mechanisms that take place during HMC production that can further facilitate the development of their production.

**Methods**

The calculations were carried out based on DFT using the Vienna ab-initio simulation package[29] where the electron–ion interactions were simulated via the projector augmented wave method.[30] The generalized gradient approximation with the of Perdew–Burke–Ernzerhof exchange-correlation function was employed.[31] The most energetically favourable MgO cluster has a cage-like configuration with $T_h$ symmetry that included six $Mg_2O_2$ rings and eight $Mg_3O_3$ to form a shortened octahedron with equivalent Mg and O vertices.[32] The system considered consisted of a MgO cluster placed in a cubic supercell with dimensions of 20 × 20 × 20 Å. A 3 × 3 × 3 k-point sampling was employed for structure optimization calculations, while a 1 × 1 × 1 k-point was used for electronic structure calculations. $Mg(OH)_2$ cluster consisting of 9 units of $Mg(OH)_2$[33] was placed in a cubic cell with dimensions of 30 × 30 × 30 Å. A 1 × 1 × 1 Å k-point sampling was applied for all optimisation and electronic structure calculations. The considered MgO and the $Mg(OH)_2$ slabs with the (001) cleaved-plane surface were selected based on the previous work.[34] A 2 × 2 × 1 Å and 1 × 1 × 1 Å k-point sampling was used for MgO and $Mg(OH)_2$ slabs, respectively.

All systems considered were totally optimized to reach atomic forces and total energies less than 0.05 eV Å$^{-1}$ and $10^{-4}$ eV, respectively. A kinetic energy cut-off of 450 eV was set for all calculations. The van der Waals–corrected functional Becke88 optimisation (optB88)[35] was adopted for the consideration of non-covalent chemical interactions between molecules and clusters. The adsorption energy of the molecule is given by the following equation[36]:

$$E_{ads} = E_{molecule/cluster} - (E_{molecule} + E_{cluster}), \qquad (1)$$

where $E_{molecule/cluster}$ is the total energy of the cluster with the adsorbed molecule, $E_{molecule}$ is the total energy of the isolated molecule and $E_{cluster}$ is the total energy of the bare cluster. Under this definition, the negative adsorption energy indicates an exothermic and favourable process. The electrons gained or lost are defined as the difference of valence electrons of an atom in the adsorbed system from the atom in a free molecule or a substrate, according to the equation $\Delta q = q_{after\ adsorption} - q_{before\ adsorption}$. The negative and positive values indicate electrons gained and lost, respectively.

The charge transfer between the molecule and the cluster is given by the charge density difference (CDD) $\Delta \rho(r)$:

$$\Delta \rho(r) = \rho_{cluster+molecule}(r) - \rho_{cluster}(r) - \rho_{mol}(r), \qquad (2)$$

where $\rho_{cluster+molecule}(r)$, $\rho_{cluster}$ and $\rho_{mol}(r)$ are the charge densities of the cluster with the adsorbed molecule, the bare cluster and the isolated molecule, respectively. The Bader analysis was used to calculate the charge transfer between the molecules and the clusters[37].

The Arrhenius equation is given by the following formula:

$$k = Ae^{-E_b/RT}, \qquad (3)$$

where $k$ is the rate constant, A is the pre-exponential factor, $E_b$ is the activation energy or the energy barrier for a reaction, R is the universal gas constant, and T is the absolute temperature.[38]

The electron localisation function (ELF) was calculated to obtain the distribution of electrons in the considered structures. The degree of charge localisation in real space is depicted by the value of the ELF (between 0 and 1), where 0 represents a free electronic state and 1 represents a perfect localisation. An isosurface value of 0.65 was adopted in this work.[39]

The climbing image–nudged elastic band (CI-NEB) method[40] was used to obtain the reaction pathway of the molecule on the cluster. The AIMD simulations were carried out at room temperature of 300 K. The simulation lasted for ~5 ps with a time step of 1 fs, and the temperature was controlled by a Nose–Hoover thermostat.[41]

**Results and Discussion**

*MgO interaction with $CO_2$ and $H_2O$.* The interaction of the MgO cluster with the $CO_2$ molecule is considered to simulate the formation of MgO–$CO_2$ (MgCO$_3$) as the main precursor to HMCs. For this, various absorption configurations of the $CO_2$ molecule on the MgO cluster are considered (more details see Figure S1 in Supporting Information). Figure 1a shows the lowest-energy configuration structure of the $CO_2$ molecule adsorbed on the MgO cluster, combined with the CDD plot. In the most stable configuration, the O atom of the $CO_2$ molecule is bonded to the Mg atom of the MgO cluster. The length of the created Mg–O bond is 2.207 Å. The length of the C–O bond of the $CO_2$ molecule is elongated from 1.174 Å (bare $CO_2$) to 1.188 Å ($CO_2$ after adsorption on MgO). It is also found that the ∠(O–C–O) angle of $CO_2$ adsorbed on the MgO cluster decreases to 171.94° compared to 179.95° for the bare $CO_2$. Table S1 (see Supporting Information) combines the results for the adsorption energy $E_{ads}$ and charge transfer $\Delta q$ between the $CO_2$ molecule and the MgO cluster. It is shown that $E_{ads}$ of the $CO_2$ molecule on the MgO cluster is -0.42 eV. According to the CDD plot (see Figure 1a), the $CO_2$ molecule acts as an acceptor to the MgO cluster with the charge transfer from the surface to the molecule of 0.092 $e$ (see Table S1 in Supporting Information), which can be attributed to the basicity of the MgO cluster, as it can donate a pair of nonbonding electrons following the Lewis base role.[21] The observed elongation of the C–O bond and the enhanced charge transfer between the cluster and molecule suggest a strong interaction between them. The high electronegativity of O atoms of the molecule can be the driving force for the observed charge transfer compared to that of Mg atoms of the cluster. However, the ELF analysis (see Figure 1b) shows that electron density is mainly located at the Mg-O bond, which indicates electron depletion from the surface of the cluster to the $CO_2$ molecule, and at the O atoms of the $CO_2$ molecule, indicating that strong covalent bonding remains only within the molecule.

To deeper understand the interaction of the $CO_2$ molecule with the MgO cluster, density of states (DOS) and local density of states (LDOS) analyses of $CO_2$-adsorbed MgO are performed (see Figure 1c). The bare MgO cluster has higher HOMO and HOMO-1 states than the $CO_2$ molecule, which indicates its tendency to oxidize the molecule, whereas the $CO_2$ molecule possess LUMO and LUMO+1 states, which verifies its ability to gain electrons. Moreover, strong overlapping of LUMO and LUMO+1 states is observed upon the interaction between the molecule and the cluster, suggesting a strong interaction between them. In addition, AIMD simulations are conducted to study the interaction of the $CO_2$ molecule with the MgO cluster at room temperature. The AIMD calculations (see Movie 1 in Supporting Information) confirm the possibility of the chemisorption of $CO_2$ on the MgO cluster at room temperature and suggests a low energy barrier $E_b$ for the reaction, as it is proposed from the $E_a$ and charge transfer calculations. Therefore, the chemisorption process of $CO_2$ on MgO is further considered.

The chemisorbed configuration of $CO_2$ is chosen based on the AIMD-obtained configuration (see Figure S2 in Supporting Information). In that case, the length of the Mg–O bond formed between the cluster and the molecule is 2.080 Å, which is shorter than that in the physisorbed state (2.207 Å). The length of the newly formed Mg–O bond in the chemisorbed configuration is 2.092 Å. The C–O bond lengths of the $CO_2$ molecule are 1.269 Å and 1.266 Å, which are significantly longer than those of the $CO_2$ in its physisorbed state (1.188 Å). This indicates that C–O bonds of $CO_2$ are highly elongated upon it interaction with Mg atoms. The ∠(O–C–O) angle of 179.95° of bare $CO_2$ decreases to 129.69° for $CO_2$ adsorbed on the MgO cluster. The CDD plot (see Figure 1d) and the Bader charge transfer analysis (see Table S1 in Supporting Information), predict that $CO_2$ is an acceptor to MgO as it accumulates 0.117 $e$ from the MgO cluster. The amount of charge transferred from MgO to chemisorbed $CO_2$ is higher than that from MgO to physiosorbed $CO_2$ (see Table S1 in Supporting Information). Furthermore, $E_{ads}$ of $CO_2$ on MgO in its chemisorbed state is -1.05 eV (see Table S1 in Supporting Information), which is more than twice higher that of $CO_2$ physiosorbed on MgO. From Figure 1e, which shows ELF of $CO_2$ chemisorbed on MgO, it is seen that electron localization located on the C–O bond formed between $CO_2$ and MgO. In addition, strong electron redistribution is observed on O atoms of $CO_2$ suggesting the formation of covalent bonds between the molecule and the cluster while the C–O covalent bonds of the $CO_2$ molecule remain stable. That contribute to the depletion of electrons from the surface to the molecule as it is observed in the CCD plot in Figure 1d.

According to the DOS and LDOS plots in Figure 1f, there is a strong hybridization of the HOMO, HOMO-1 and LUMO+1 states of the MgO cluster and the $CO_2$ molecule, indicating a strong interaction between them and signifying the possibility of chemisorption of the $CO_2$ molecule on the MgO cluster. The AIMD simulations also suggest that the chemisorption of $CO_2$ on MgO is favourable (see Movie 1 and Figure S2 in Supporting Information). Thus, the possible reaction mechanism for the transformation process for the $CO_2$ molecule on the MgO cluster from physiosorbed to chemisorbed state is further studied through NEB approach. The energy profile and related atomic configurations for the initial state (IS), transition state (TS), intermediate states (IM) and final state (FS), showing the transition of the $CO_2$ molecule from the physiosorbed state to the chemisorbed state, are depicted in Figure 1g. TS with an energy level of 0.049 eV proposes the low energy barrier $E_b$ for this transition (see Table S2 in Supporting Information). It seems that the O atom of $CO_2$ has a high tendency to oxidize the Mg atom of the cluster. This oxidation is expedited at IM3 by the approach and further bonding of the C atom of the molecule to the O of the cluster, which leads to a drop of $E_b$ to 0.001 eV. At FS, the second O atom of $CO_2$ is bonded to the Mg atom of the cluster, and $E_b$ further drops to -0.617 eV which suggests the reaction is exothermic.

To summarize, the elongation of the C–O bond and the decrease of the ∠(O–C–O) angle of the $CO_2$ molecule upon its chemisorption on the MgO cluster comparing to physisorption lead to an increase of $E_a$.[24] In addition, higher charge transfer from the cluster to the $CO_2$ molecule during chemisorption stabilizes the adsorption of the $CO_2$ molecule on the cluster[42]. These results are well agreed with found low $E_b$ of exothermic transition of $CO_2$ from physisorbed state to chemisorbed state and with experimental observations confirming that the calcination of magnesite ($MgCO_3$) is an endothermic process.[5] Therefore, the chemisorption of $CO_2$ on MgO occurs favourably under the reaction conditions.

The reaction of $H_2O$ with MgO leads to the formation of $Mg(OH)_2$, a phase that might also undergo carbonation, which results in the HMC formation. Hence, the hydration of the MgO cluster is also investigated. All possible absorption configurations of $H_2O$ on the MgO cluster are considered (see Figure S3 in Supporting Information). According to Table S1 (see Supporting Information), $E_{ads}$ for the most energetically favourable configuration of adsorbed $H_2O$ on the MgO cluster (see Figure 2a) is -0.95 eV. In this configuration, the Mg–O bond between the O atom of the $H_2O$ molecule and the Mg atom of the MgO cluster and the H–O bond between the H atom of the molecule and the O atom of the cluster are formed. The length of the Mg–O and H–O bonds is found to be 2.085 Å and 1.627 Å, respectively. Moreover, the length of the H–O bond of the $H_2O$ molecule before bonding to the cluster is 0.936 Å, and it is elongated to 1.036 Å after adsorption, signifying the tendency of $H_2O$ to bind to MgO. According to the Bader charge transfer analysis and the CCD plot (see Figure 2a) the $H_2O$ molecule is a strong electron acceptor to the cluster with $\Delta q = -1.122$ $e$ (see Table S1 in Supporting Information). The basicity of the MgO cluster facilitates the electrons transfer from the O atom of the cluster to the $H_2O$ molecule, while a higher electronegativity of the O atom of the molecule facilitates electron depletion towards H atoms. Such significant charge redistribution between the MgO cluster and the H$_2$O contributes to its adsorption.[36]

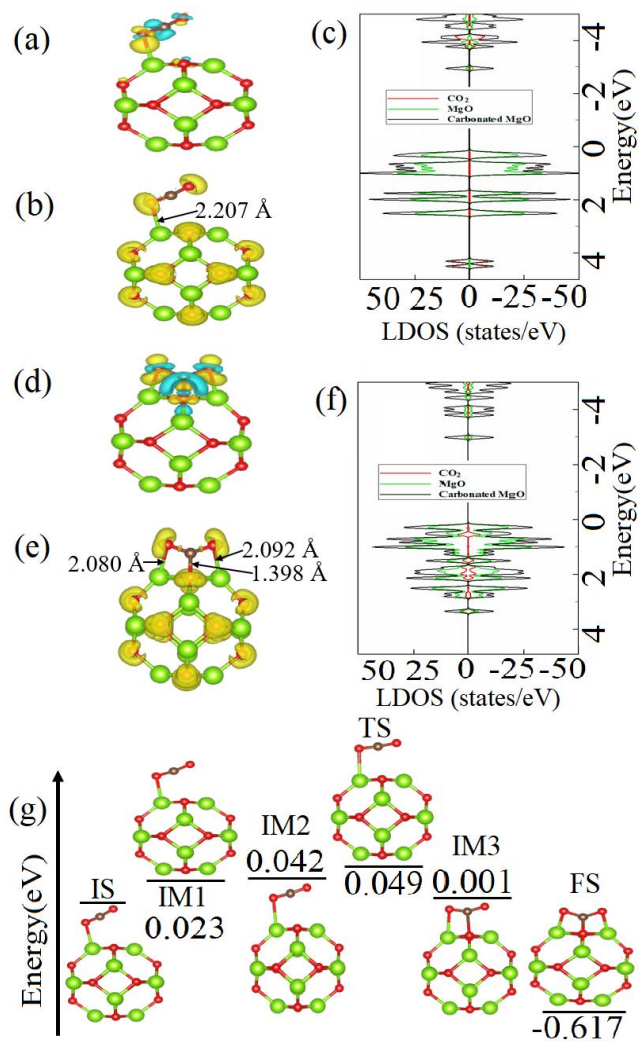

**Figure 1.** (a) The lowest-energy configuration of the CO$_2$ molecule physisorbed on the MgO cluster combined with the CDD isosurface plot (0.003 Å$^{-3}$). (b) The ELF and (c) DOS and LDOS for the CO$_2$–physiosorbed MgO cluster. (d) The lowest-energy configuration of the CO$_2$ molecule chemisorbed on the MgO cluster combined with the CDD isosurface plot (0.009 Å$^{-3}$). (e) The ELF and (f) DOS and LDOS for the CO$_2$–chemisorbed MgO cluster. (g) The energy barrier and atomic structures corresponding to the minimum energy pathway for the chemisorption process of the CO$_2$ molecule on the MgO cluster.

The ELF plot in Figure 2b shows the electron localization between the O atom of the H$_2$O molecule and the Mg atom of the MgO cluster, as well as the localization between the H atom of the H$_2$O molecule and the O atom of the MgO cluster, which confirms electron depletions at these sites and suggests the formation of the H-O and Mg-O bonds between the molecule and the cluster. The DOS and LDOS plots in Figure 2c display the hybridization of H$_2$O and MgO states at -3.3 eV and 4.9 eV and a weak interaction at 4.4 eV. The conducted AIMD simulation also confirms the dissociation of the H$_2$O molecule on the MgO cluster and formation of the H-O and Mg-O bonds (see Movie 2 and Figure S4 in Supporting Information).

NEB calculations are carried out to show the possible reaction mechanism of the H$_2$O molecule dissociation on the MgO cluster. Figure 2d presents the energy profile and related atomic configurations for the IS, TS, IMs and FS showing the dissociation of the H$_2$O molecule of the MgO cluster. As it is seen, between IS and TS the H$_2$O molecule bonding to the MgO cluster through the rotation of the H atom of the molecule (IM1). $E_b$ of the H$_2$O molecule dissociation on the MgO cluster at TS is found to be as high as 0.245 eV (see Table S2 in Supporting Information). Further reaction at IM2 and IM3 leads to the bonding of the H atom of the H$_2$O molecule to the nearest O atom of the cluster and the consequent H$_2$O dissociation at FS occurring with an energy release of 0.179 eV. A higher energy release during the carbonation (-0.617 eV) of the MgO cluster compared to that during hydration (-0.179 eV) of the MgO cluster, indicates that the carbonation of the MgO cluster is a more exothermic process than its hydration. Therefore, the carbonated MgO is more thermodynamically stable. However, $E_b$ for carbonation of the MgO cluster is 0.235 eV, which is lower than $E_b$ of 0.245 eV for hydration of the MgO cluster. On the other hand, AIMD simulations suggest that the hydration of the MgO cluster passes faster than its carbonation (see Figure S4 in Supporting Information). Therefore, hydration and carbonation rates of the MgO cluster are compared based on the Arrhenius equation (Eq. 3), according to which the reaction rate depends on two factors: activation energy of the reaction and pre-exponential factor A. Therefore, besides the calculated $E_b$, the A factor, describing the frequency of collisions between reactant molecules at a standard concentration, should be taken into consideration for the comparison of hydration and carbonation rates of the MgO cluster. The hydrolysis of the MgO cluster changes its structure due to a break of Mg-O bonds of the MgO cluster upon interaction with H$_2$O, while the carbonation of the MgO cluster does not cause the alteration of the MgO cluster. This leads to a significant difference in the A factor for the hydration and carbonation of the cluster. As a result, the hydration of the MgO cluster is faster than its carbonation as it is shown by AIMD simulations (see Figure S4 and Movies 1 and 2 in Supporting Information). This observation is also in line with the fact that $E_{ads}$ of the H$_2$O molecule (-0.95 eV) on the MgO cluster is more than 2 times lower than that of the CO$_2$ molecule (-0.42 eV) on the MgO cluster, which leads to faster hydration reaction. Faster hydration of the MgO cluster is also observed in AIMD simulations (see Figures S4 in Supporting Information) where the adsorption of the H$_2$O molecule of the MgO cluster occurs ~3 times faster than that of the CO$_2$ molecule. Furthermore, to compare the hydration and the carbonation rate of the MgO cluster, AIMD simulations are performed to simulate a CO$_2$- and H$_2$O-saturated environment, consisting of three CO$_2$ and three H$_2$O molecules (see Movie 3 in Supporting Information). The trajectory of these molecules shows that hydration of MgO is significantly faster than its carbonation (see Figure S5 in Supporting Information), as all three considered H$_2$O molecules bond to the MgO cluster before any of the CO$_2$ molecules.

In summary, the formation of the H-O and Mg-O bonds between the H$_2$O molecule and the MgO cluster verifies H$_2$O chemisorption on the cluster. The calculated NEB energy profile diagram predicts that the H$_2$O molecule dissociation on the MgO cluster is an exothermic process, and the carbonation of MgO is thermodynamically more favourable than its hydration. However, although the calculated $E_b$ for the hydration of the MgO cluster is higher than that for its carbonation, the hydration of the MgO cluster is found to be faster, as confirmed by the calculated $E_{ads}$ and AIMD simulations.

As it is found that hydration of MgO occurs faster than it carbonation, the CO$_2$ molecule interaction with the hydrated MgO cluster (previously found lowest-energy configuration of hydrated MgO is used) is studied. Several possible configurations of the CO$_2$ molecule on the hydrated MgO cluster are considered (see Figure S6 in Supporting Information). Figure 3a shows the lowest-energy configuration of the CO$_2$ molecule on the hydrated MgO cluster, where the O atom of the CO$_2$ molecule is bonded to the Mg atom of the MgO cluster. The newly formed Mg–O bond has a length of 2.175 Å. The length of the C–O bond (the one closest to the cluster) of the adsorbed CO$_2$ is elongated to 1.182 Å compared to that of bare CO$_2$ of 1.174 Å, while another C–O bond of CO$_2$ shortens to 1.165 Å. The ∠(O–C–O) angle of the $CO_2$ molecule also decreases from 179.95° to 174.43° upon its adsorption.

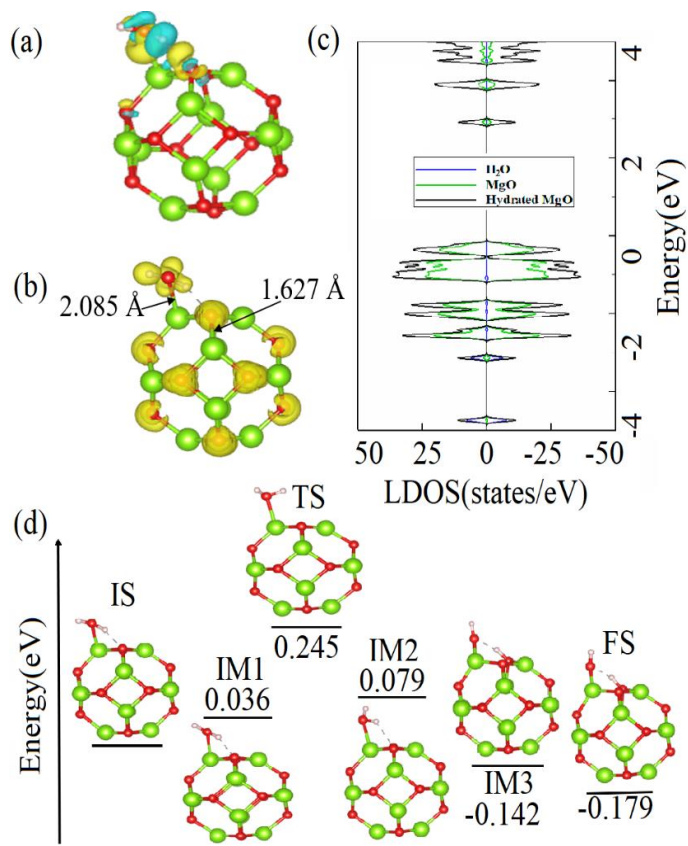

**Figure 2.** (a) The lowest-energy configuration of the $H_2O$ molecule on the MgO cluster combined with the CDD isosurface plot (0.003 Å-3). (b) The ELF and (c) DOS and LDOS for the $H_2O$–adsorbed MgO cluster. (d) The energy barrier and atomic structures corresponding to the minimum energy pathway for the hydration of the MgO cluster.

The CDD plot in Figure 3a shows that the $CO_2$ molecule is an acceptor to the hydrated MgO cluster as there is a depletion of the electron on the Mg atom of the cluster and accumulation of electrons on the O atom of the $CO_2$ molecule. The Bader charge transfer analysis predicts that the amount of the charge transferred from the cluster to the molecule is 0.058 $e$. Importantly, $E_{ads}$ of the $CO_2$ molecule on the hydrated MgO cluster is -0.53 eV, which is lower than that of the $CO_2$ molecule on the bare MgO cluster. This suggests stronger bonding of the $CO_2$ molecule with the hydrated MgO cluster compared to the bare MgO cluster. The ELF plot in Figure 3b demonstrates the electron localization between the O atom of the molecule and the Mg atom of the cluster. It verifies the accumulation of electrons on the O atoms of the $CO_2$ molecule and suggests that the C–O bonds of the molecule remain covalent. The DOS and LDOS plots in Figure 3c show strong overlapping of HOMO states of the $H_2O$ molecule and the MgO cluster in the range from -4.2 to -4.8 eV and strong overlapping of LUMO+1 states of the $H_2O$ molecule and the MgO cluster in the range from 3.4 to 4.5 eV, which confirms a strong bonding between the $CO_2$ molecule and the hydrated MgO cluster. The conducted AIMD calculations predict the possibility of chemisorption of the $CO_2$ molecule on the hydrated MgO cluster at room temperature. In chemisorbed state, the O atoms of the $CO_2$ molecule are bonded to the Mg atoms of the hydrated MgO cluster and the C atom of the $CO_2$ molecule is bonded to the O atom of the hydrated MgO cluster (see Movie 4 and Figure S7 in Supporting Information). It is observed that the carbonation of bare MgO occurs slower than the carbonation of hydrated MgO due to the formation of OH groups on the MgO cluster during its hydration, that hinder the carbonation process.

To gain insights into the carbonation mechanism of hydrated MgO, the chemisorption process of $CO_2$ on it is considered. The lowest-energy configuration of chemisorbed $CO_2$ molecule on the hydrated MgO cluster (for more details see Figure S7 in Supporting Information) is shown in Figure 3d. Here, both O atoms of the $CO_2$ molecule form chemical bonds with the Mg atoms of the hydrated MgO cluster. The C–O bonds of the $CO_2$ molecule are elongated to 1.269 Å and 1.275 Å (compared to 1.174 Å of the bare $CO_2$ molecule) upon it adsorption on the hydrated MgO cluster. The length of newly formed Mg-O bonds is 2.057 Å and 2.025 Å, while the length of the C–O bond formed between the C atom of the molecule and the O atom of the cluster is 1.382 Å. The $\angle$(O–C–O) angle of the adsorbed $CO_2$ molecule is found to be 129.16°, which is lower than that of the $CO_2$ molecule in its physisorbed state. The CDD plot in Figure 3d shows that the charge is mostly distributed on the $CO_2$ molecule and partially on the O atom of the MgO cluster bonded to the C atom of the $CO_2$ molecule. The basicity of the hydrated MgO cluster drives the electron transfer from the molecule to the hydrated cluster. According to the Bader charge transfer analysis, the chemisorbed $CO_2$ molecule gains 0.086 $e$ from the hydrated MgO cluster. Therefore, the amount of the charge transferred from the hydrated MgO cluster to the chemisorbed $CO_2$ molecule is higher than that from the hydrated MgO cluster to the physisorbed $CO_2$ molecule (see Table S1 Supporting Information). The calculated $E_{ads}$ of -1.55 eV for the $CO_2$ molecule chemisorbed on the hydrated MgO cluster is higher than that of the $CO_2$ molecule chemisorbed on the bare MgO cluster (-1.05 eV). The ELF plot in Figure 3e depicts electron localizations between the O atoms of the chemisorbed $CO_2$ molecule and the Mg atoms of the hydrated MgO cluster and the C atom of the chemisorbed $CO_2$ molecule and the O atom of the hydrated MgO cluster, which suggests the existence of the covalent Mg-O and C-O bonds between the cluster and the molecule. Meanwhile, the covalent bonding between the $H_2O$ molecule and the MgO cluster remains unchanged. According to DOS and LDOS plots presented in Figure 3f, there is a strong hybridization of HOMO and HOMO-1 states of the hydrated MgO cluster and the chemisorbed $CO_2$ molecule. The overlapping of the cluster and the molecule is also observed at -3.8 eV, -4.5 eV and 4.6 eV.

$E_b$ of 0.275 eV (see Table S2 in Supporting Information) for the transition of the $CO_2$ molecule from physisorbed state to chemisorbed state on the hydrated MgO cluster is calculated by the NEB approach (see Figure 3g). The transition involves the IM2 stage, where the O atom of the $CO_2$ molecule oxidizes the Mg atom of the MgO cluster, which leads to the drop of $E_b$ to 0.153 eV. This triggers an exothermic process of bonding the C and O atoms of the $CO_2$ molecule to the hydrated MgO cluster at the FS state via the IM3 (-0.800) state with the released energy of 1.028 eV. According to calculated reaction energies in the carbonation process of bare MgO (-0.617 eV) and hydrated MgO (-1.028 eV), carbonation of the hydrated MgO is thermodynamically more favourable. However, $E_b$ for the transition of $CO_2$ from the physisorbed state to the chemisorbed state on hydrated MgO (0.275 eV) is higher than that of $CO_2$ on bare MgO (0.234 eV). Therefore, $CO_2$ chemisorption on hydrated MgO is kinetically unfavourable. This matches the AIMD simulation results (see Figures S2 and S6 in Supporting Information), where the carbonation of bare MgO is faster than that of the hydrated MgO. Importantly, this verifies the fact that the initial hydration of MgO can hinder its carbonation.[15]

In summary, the chemisorption of the $CO_2$ on the MgO cluster is found to be the most energetically favourable. The charge redistribution between the MgO cluster and the $CO_2$ molecule during the chemisorption[37,38] and the comparison of the energy released during carbonation of the bare and the hydrated MgO clusters suggests carbonation of the bare MgO cluster is faster than that of the hydrated MgO cluster, which uncovers the hindrance effect of $H_2O$ on the carbonation of MgO.

The observed results are also supported by AIMD simulations (see Movie 4 and Figure S7 in Supporting Information).

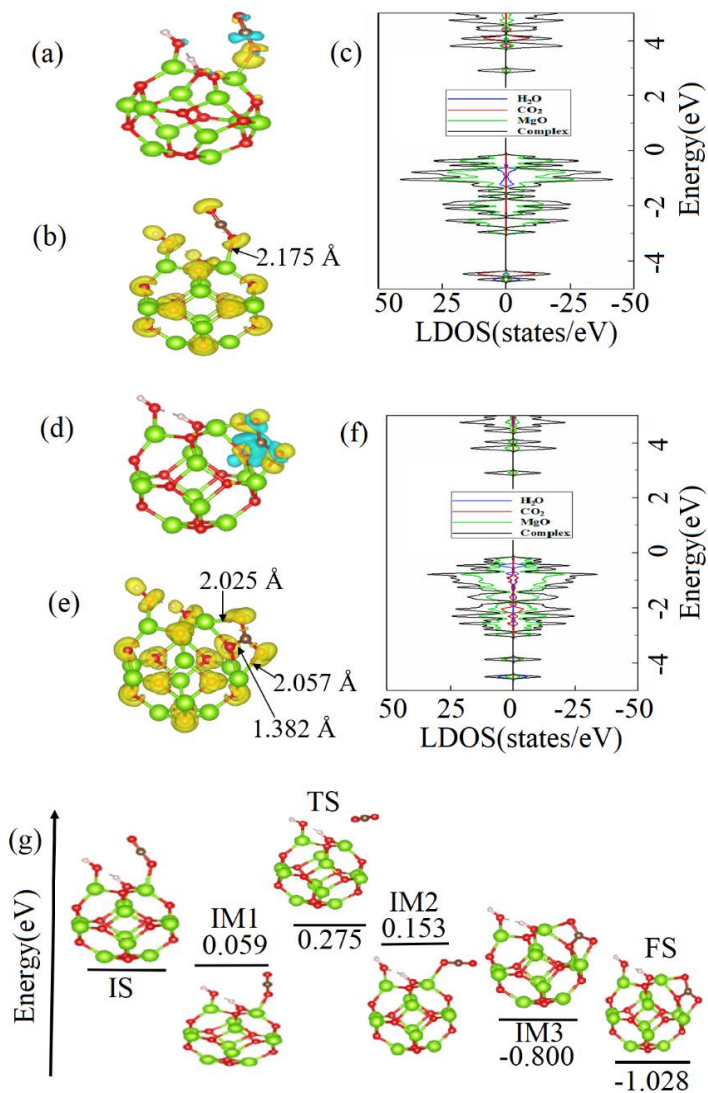

**Figure 3.** (a) The lowest-energy configuration of the physisorbed $CO_2$ molecule on the hydrated MgO cluster combined with the CDD isosurface plot (0.003 Å$^{-3}$). (b) The ELF and (c) DOS and LDOS for the $CO_2$–physiosorbed hydrated MgO cluster. (d) The lowest-energy configuration of the chemisorbed $CO_2$ molecule on the hydrated MgO cluster combined with the CDD isosurface plot (0.009 Å$^{-3}$). (e) The ELF and (f) DOS and LDOS for the $CO_2$–chemisorbed hydrated MgO cluster. (g) The energy barrier and atomic structures corresponding to the minimum energy pathway for the transition of the the $CO_2$ molecule from physiorbed to chemisorbed states on the hydrated MgO cluster.

*$Mg(OH)_2$ interaction with $CO_2$ and $H_2O$.* In RMC reactions, the carbonation of $Mg(OH)_2$ leads to the production of a range of HMCs.[12,13] Therefore, the mechanism of the carbonation of $Mg(OH)_2$ is further studied. For that, several possible configurations of the $CO_2$ molecule and the $Mg(OH)_2$ cluster are examined (see Figure S8 in Supporting Information). The most favourable sites for the $CO_2$ molecule adsorption on $Mg(OH)_2$ are located at its edges. Figure 4a combines the atomic structure of the lowest-energy configuration and the CDD plot for the $CO_2$ molecule adsorbed on the $Mg(OH)_2$ cluster. In that case, the C atom of the $CO_2$ molecule is located below the O atom at the edge of the $Mg(OH)_2$ cluster and is bonded to the O atom of the cluster. In the same way, the O atom of the $CO_2$ molecule is bonded to the Mg atom at the edge of the $Mg(OH)_2$ cluster and forms the Mg–O bond of a length of 2.069 Å. Upon adsorption, the C–O bonds in the $CO_2$ molecule elongates from 1.174 Å (bare CO$_2$) to 1.266 Å, while a newly formed C–O bond between the CO$_2$ molecule and Mg(OH)$_2$ has a length of 1.515 Å and ∠(O–C–O) changes from 179.95° to 136.88°. The CDD plot in Figure 4a displays the charge transfer from O atoms at the edge of Mg(OH)$_2$ to the CO$_2$ molecule. The Bader charge transfer analysis suggests that CO$_2$ acts as an acceptor to the Mg(OH)$_2$ cluster, with the charge transfer from the cluster to the molecule of 0.397 $e$ (see Table S1 in Supporting Information). This verifies the Lewis basicity of the Mg(OH)$_2$ cluster. According to Table S1 (see Supporting Information), the $E_{ads}$ of CO$_2$ on Mg(OH)$_2$ is -0.69 eV.

The ELF plot in Figure 4b shows electron localization between the O atom of the CO$_2$ molecule and the Mg atom of the Mg(OH)$_2$ cluster, which characterizes electron transfer and strong bonding between the molecule and the edge of the cluster. The covalent bonding within the molecule also remains stable, as predicted by the charge localization on both the C–O bonds of the CO$_2$ molecule. The DOS and LDOS plots for the CO$_2$–adsorbed Mg(OH)$_2$ cluster are shown in Figure 4c. The observed strong orbital hybridisation of CO$_2$ and Mg(OH)$_2$ at the energy of -1.7 eV and in a range from -2 to -3.7 eV confirms the strong interaction between CO$_2$ and Mg(OH)$_2$ proposed by the charge transfer and ELF analysis. Figure 4d depicts the potential energy profile and atomic structures corresponding to the minimum energy pathway for the carbonation of the Mg(OH)$_2$ cluster. It is shown that $E_b$ for the carbonation of Mg(OH)$_2$ is as low as 0.002 eV (TS in Figure 4d), which is equivalent to a spontaneous process at room temperature. To reach the chemisorbed state at FS (-0.303 eV), the CO$_2$ molecule passes through the IM2 state (-0.064 eV) where the C atom of the molecule bonds to the O atom of the cluster, and IM3 (-0.256), at which point the O atom of the molecule forms a bond with the Mg atom of the cluster. It is also noted that the carbonation of Mg(OH)$_2$ is a highly exothermic process.

In summary, the elongation of C–O bonds and the decrease of ∠(O–C–O) of the CO$_2$ molecule, along with the strong charge transfer between the molecule and the Mg(OH)$_2$ cluster, play a dominant role in CO$_2$ chemisorption on Mg(OH)$_2$. Despite the chemisorption of CO$_2$ on Mg(OH)$_2$ occurring only at the edges, chemisorption mechanism of CO$_2$ for Mg(OH)$_2$ is similar to that for MgO. In both cases, chemisorption of CO$_2$ is an exothermic process with low $E_b$ and significant energy release. However, the activation energy for the carbonation of Mg(OH)$_2$ (0.002 eV) is significantly lower than that for MgO (0.049 eV), confirming that carbonation of the Mg(OH)$_2$ is faster than that of the MgO. In turn, the energy released during the MgO carbonation (-0.617 eV) is about 2 times lower than that during the Mg(OH)$_2$ carbonation (-0.303 eV), which suggests that the carbonation of the MgO cluster is more thermodynamically favourable.

Mg(OH)$_2$ is often affected by aqueous environments, therefore, the interaction of the H$_2$O molecule with the Mg(OH)$_2$ cluster can play a key role in HMC formation. From the studied configurations for that interaction of H$_2$O with Mg(OH)$_2$ (see Figure S9 in Supporting Information), the lowest-energy configuration is related to the H$_2$O molecule located at the edge of the Mg(OH)$_2$ cluster. The length of the O–H bond of the bare H$_2$O molecule (1.972 Å) is shortened to 1.020 Å upon the H$_2$O molecule bonding to the Mg(OH)$_2$ cluster, while the O-H bond at the edge of the Mg(OH)$_2$ cluster is elongated from 0.965 Å to 0.984 Å. The CDD plot in Figure 4e shows that the charge is mostly distributed on the H$_2$O molecule and partially at the edge of the Mg(OH)$_2$ cluster. The Bader charge transfer analysis predicts the H$_2$O molecule to be a weak acceptor to the Mg(OH)$_2$ cluster that accumulates 0.044 $e$ (see Table S1 in Supporting Information). $E_{ads}$ of the H$_2$O molecule on the Mg(OH)$_2$ cluster is -0.74 eV (see Table S1 in Supporting Information). The ELF plot in Figure 4f shows insignificant electron distributions between the O atom of the H$_2$O molecule and the H atom of the Mg(OH)$_2$ cluster. Meanwhile, low electron density between the H atom of the H$_2$O molecule and the O atom of the Mg(OH)$_2$ cluster indicates weak interaction between them. Moreover, orbital localization between the O–H bonds of the H$_2$O molecule shows that the covalent bonds of the molecule remain stable. The DOS and LDOS plots for the H$_2$O molecule adsorbed on the Mg(OH)$_2$ cluster, shown in Figure 4g, also suggests a weak interaction between the molecule and the cluster at -1.5 eV, -1.8 eV, -2 eV, and in ranges from -2.2 to -2.5 eV and from -4.1 to -4.2 eV.

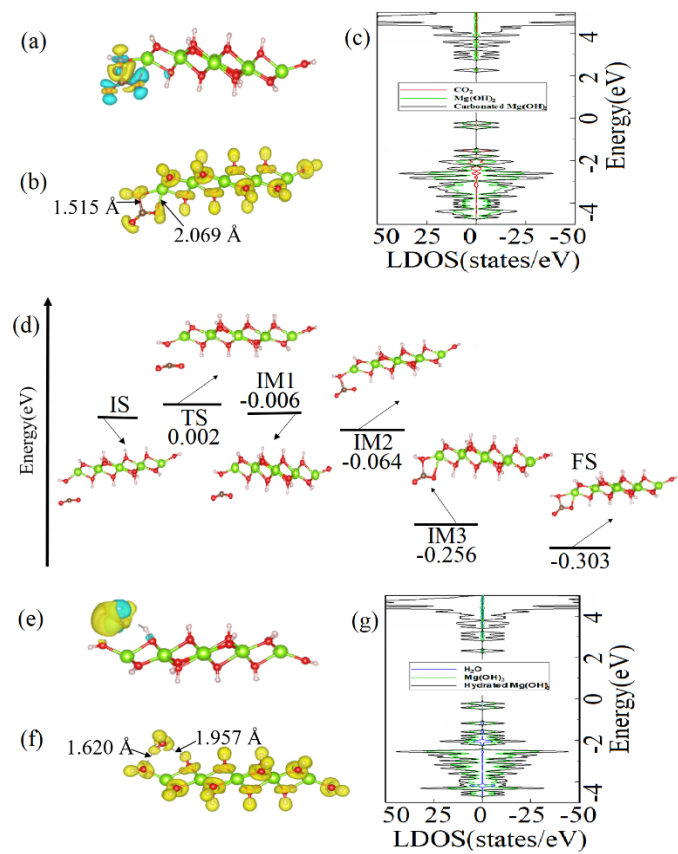

**Figure 4.** (a) The atomic structure of the lowest-energy configuration of the CO$_2$ molecule on the Mg(OH)$_2$ cluster combined with the CDD isosurface plot (0.006 Å$^{-3}$). (b) The ELF and (c) DOS and LDOS for the CO$_2$–adsorbed Mg(OH)$_2$ cluster. (d) The energy barrier and atomic structures corresponding to the minimum energy pathway for the carbonation of the Mg(OH)$_2$ cluster. (e) The atomic structure of the lowest-energy configuration of the H$_2$O molecule on the Mg(OH)$_2$ cluster combined with the CDD isosurface plot (0.009 Å$^{-3}$). (f) The ELF and (g) DOS and LDOS for the H$_2$O–adsorbed Mg(OH)$_2$ cluster.

In summary, it is found that the H$_2$O molecule is located at the edges of the Mg(OH)$_2$ cluster. The calculated low $E_{ads}$ and weak charge transfer between the H$_2$O molecule and the Mg(OH)$_2$ cluster suggest that H$_2$O is physisorbed on Mg(OH)$_2$. However, it is well-known that the presence of H$_2$O facilitates the formation of HMCs in accessible pores during the carbonation process.[37]

To investigate a mechanism of reaction of nesquehonite formation (Mg(OH)$_2$ + CO$_2$ + 2H$_2$O → MgCO$_3$·3H$_2$O), at the first step, the simultaneous interaction of the carbonated Mg(OH)$_2$ cluster and the H$_2$O molecule (Mg(OH)$_2$ + CO$_2$ + H$_2$O) is considered. At the second step, one more H$_2$O molecule is introduced to the system studied at the first step (CO$_2$+ 2H$_2$O+ Mg(OH)$_2$). Although the natural process of the nesquehonite formation also includes nucleation and growth from species in solution, the studied reaction will still help to understand possible nucleation or growth paths of nesquehonite. At the first step, various configurations of one H$_2$O molecule (see Figure S10 in Supporting Information) on the carbonated Mg(OH)$_2$ cluster are considered. At the lowest-energy configuration, the H$_2$O molecule is bonded to the edge of the Mg(OH)$_2$ cluster. The length of the O–H bonds increases to 0.970 Å and 1.022 Å compared to these of the bare $H_2O$ molecule (0.936 Å). The distance between the H atom of the $H_2O$ molecule and the O atom of the $Mg(OH)_2$ cluster is 1.620 Å, while the distance between the O atom of the $H_2O$ and the H atom of the $Mg(OH)_2$ cluster is 1.957 Å. According to Table S1 (see Supporting Information), $E_{ads}$ of the $H_2O$ molecule on the $Mg(OH)_2$ cluster is -0.86 eV. The CDD plot in Figure 5a shows that there is a depletion of electrons at the edge O atoms of the $Mg(OH)_2$ cluster and charge accumulation at the H atoms of the $H_2O$ molecule. The Bader charge transfer analysis shows that the $H_2O$ molecule gains 0.046 $e$ from the $Mg(OH)_2$ cluster which confirms that $H_2O$ is a weak acceptor to $Mg(OH)_2$ (see Table S1 in Supporting Information). Further, the ELF plot in Figure 5b shows $H_2O$ is physisorbed on $Mg(OH)_2$, as there is no electron density localization between the $Mg(OH)_2$ cluster and the $H_2O$ molecule, while the H-O bonds of $H_2O$ remain their covalent nature. Figure 5c represents the DOS and LDOS plots for the $H_2O$ molecule adsorption on the carbonated $Mg(OH)_2$ cluster. Small overlapping of states of the $H_2O$ molecule and the carbonated $Mg(OH)_2$ cluster is observed in the range from -3.0 to -3.7 eV, verifying weak interaction between them.

Further, at the second step, the second $H_2O$ molecule is introduced to the $Mg(OH)_2 + CO_2 + H_2O$ system obtained at the first step (see Figure S11 in Supporting Information). The lowest energy configuration of the $H_2O$ molecule on the $Mg(OH)_2 + CO_2 + H_2O$ system is shown in Figure 5d, where the second $H_2O$ molecule is also located at the edge of the carbonated $Mg(OH)_2$ cluster. The length of the H-O bonds of the bare $H_2O$ molecule is 0.971 Å, while they slightly elongate to 0.975 Å and 0.973 Å after adsorption. The distance between the O atom of the $H_2O$ molecule and the Mg atom of the $Mg(OH)_2$ cluster is 2.232 Å. According to Table S1 (see Supporting Information) $E_{ads}$ of the $H_2O$ molecule on the $Mg(OH)_2 + CO_2 + 2H_2O$ system is -0.46 eV. The CDD plot in Figure 5d displays the depletion of electrons at O atoms located at the edge of the $Mg(OH)_2$ cluster and charge accumulation at the H atoms of the $H_2O$ molecule. According to the Bader charge transfer analysis (see Table S1 in Supporting Information) the $H_2O$ molecule is a weak acceptor to the $Mg(OH)_2$ cluster with the charge transfer of 0.037 $e$ from the cluster to the molecule. The ELF plot in Figure 5e shows no election density localization between the second $H_2O$ molecule and the $Mg(OH)_2 + CO_2 + H_2O$ system, which means there is a weak interaction between them. The DOS and LDOS plots in Figure 5f also display a weak interaction of the $H_2O$ molecule and the cluster at the range from -2.8 to -3.5 eV.

AIMD simulations are used to investigate the reaction for the formation of HMCs via the interaction of the $CO_2$ and $H_2O$ molecules with the $Mg(OH)_2$ cluster (see Movie 5 and Figures 5g and h in Supporting Information,). As it is shown, the $CO_2$ and $H_2O$ molecules are bonded at the edges of the $Mg(OH)_2$ cluster, which suggests that the formation of HMCs starts at the edges of $Mg(OH)_2$. Further, AIMD simulations are conducted to consider the effect of large amount of $H_2O$ and $CO_2$ molecules on the formation of HMCs (see Movie 6 in Supporting Information). For that, two $H_2O$ and one $CO_2$ molecules are added to the previously considered $Mg(OH)_2 + CO_2 + 2H_2O$ system. As shown in Figure 5h, the first $CO_2$ molecule is able to carbonate the $Mg(OH)_2$ cluster. However, after the bonding of the first $CO_2$ molecule to the cluster, the second $CO_2$ molecule is unable to bind to the carbonated $Mg(OH)_2$ cluster (Figures 5i and j). This suggests that the formation of an early layer of carbonates in RMC-based concrete formulations may limit the continuation of carbonation by forming a physical barrier that prohibits further interaction between $Mg(OH)_2$ and $CO_2$. These limitations in carbonation of $Mg(OH)_2$ can cause large amounts of unreacted crystals leading to relatively low strength and porous microstructures.[15,43] Although the presence of $H_2O$ molecules provides the medium for carbonation and further transformation of $Mg(OH)_2$ into HMCs and is required for the continuous formation of HMCs[44], according to the AIMD simulations, excessive $H_2O$ hinders $CO_2$ penetration to the $Mg(OH)_2$ surface. Therefore, to maintain $CO_2$ diffusion for carbonation of Mg(OH)$_2$, the amount of H$_2$O should be properly controlled.[40] The predicted results of Mg(OH)$_2$ passivation with the formation of the barrier of carbonates and H$_2$O hindrance effect on carbonation of MgO correspond to the carbonation mechanisms of portlandite.[45]

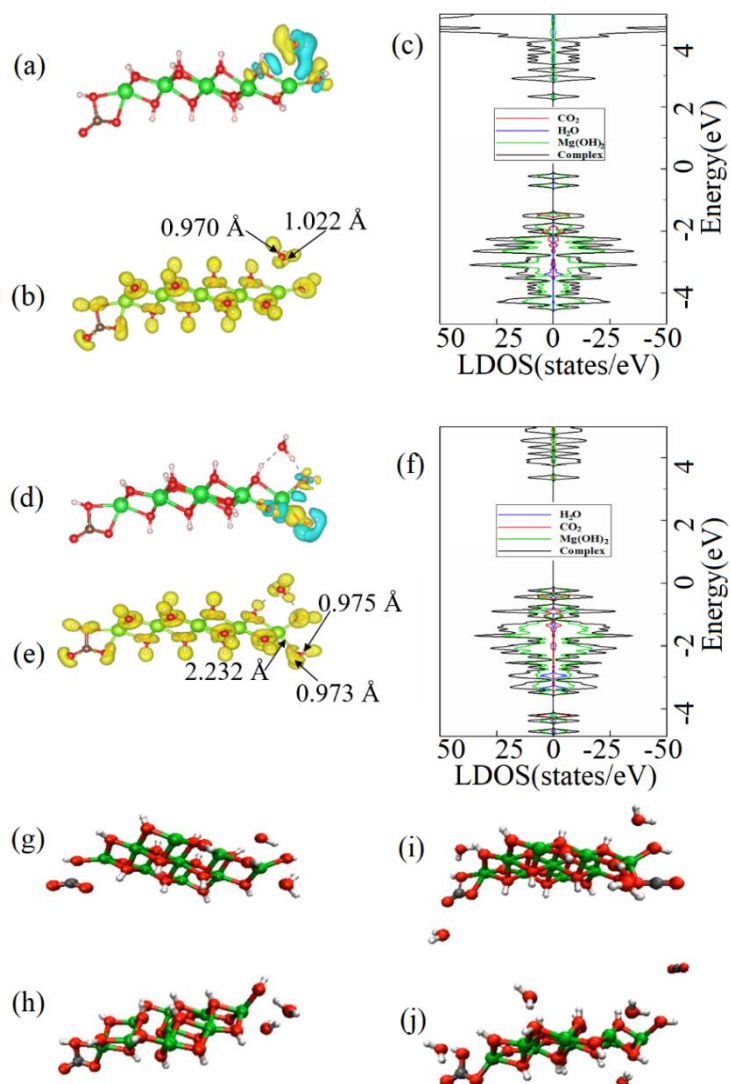

**Figure 5.** (a) The lowest-energy configuration of the H$_2$O molecule on the carbonated Mg(OH)$_2$ cluster with the CDD isosurface plot (0.001 Å$^{-3}$). (b) The ELF and (c) DOS and LDOS for H$_2$O molecules on the carbonated Mg(OH)$_2$ cluster. (d) The lowest-energy structure of the H$_2$O molecules on the carbonated Mg(OH)$_2$ cluster combined with the CDD plot (0.001 Å$^{-3}$). (e) The ELF and (f) DOS and LDOS for the H$_2$O molecule on the carbonated Mg(OH)$_2$ cluster. (g) Initial and (h) final (chemisorbed state) configurations of CO$_2$ and H$_2$O molecules on the Mg(OH)$_2$ cluster. (i) Initial and (h) final configurations of CO$_2$ and H$_2$O molecules on the carbonated Mg(OH)$_2$ cluster.

As the adsorption behaviour of molecules on the clusters may be different from that on the bulk materials. Further, the interaction of the CO$_2$ and H$_2$O molecules with clusters and with the bulk MgO and Mg(OH)$_2$ is compared. Figures 6a and b represent the lowest-energy configuration of the CO$_2$ and H$_2$O molecules on MgO(001). Figure 6 a shows the O atom of CO$_2$ molecule is located above Mg atom of MgO(001). The length of C–O bonds in the CO$_2$ molecule elongates from 1.174 Å (bare CO$_2$) to 1.182 Å and 1.175 Å and ∠(O–C–O) decreases from 179.95° to 176.81°. The calculated $E_{ads}$ for CO$_2$ on the MgO(001) surface is -0.34 eV (see Table S3 in Supporting Information). This indicates weak adsorption of CO$_2$ on the bulk MgO(001) compared to the physisorbed-CO$_2$ on the MgO cluster ($E_{ads}$ = -0.42 eV). According to Figure 6b, O atom of H$_2$O molecule is located above Mg atom of MgO(001). The length of the H–O bonds of the H$_2$O molecule, elongates from 0.972 Å (bare H$_2$O) to 0.983 Å and 0.977 Å. $E_{ads}$ = -0.58 eV of the H$_2$O molecule on MgO(001) (see Table S3 in Supporting Information) is found to be lower than that of the H$_2$O molecule (-0.95 eV) on the MgO cluster. The length of the Mg–O bond formed between the O atom of the H$_2$O molecule and the Mg atom at the MgO(001) surface is 2.239 Å, which is longer than the Mg–O bond formed between the O atom of the H$_2$O molecule and the Mg atom of the MgO cluster (2.085 Å). Therefore, for both H$_2$O and CO$_2$ molecules their adsorption and possible dissociation at the edges or defective surfaces of the MgO crystal, presented here via clusters, is more favourable. This result well matches the previously reported observation on weak adsorption of the CO$_2$ molecule on the MgO(001) surface.[24]

Figures 6c and d show the lowest-energy configurations of the CO$_2$ and H$_2$O molecules on the Mg(OH)$_2$(001) surface. Figure 6c indicates the location of the C atom of the CO$_2$ molecule is located above the Mg-O bond of the Mg(OH)$_2$(001) surface. The C–O bonds in the CO$_2$ molecule elongates from 1.174 Å (bare CO$_2$) to 1.177 Å and 1.178 Å, and ∠(O–C–O) changes from 179.95° to 179.00°. $E_{ads}$ of the CO$_2$ molecule on the Mg(OH)$_2$(001) surface is as low as -0.25 eV, which is significantly lower than that of the CO$_2$ molecule on the Mg(OH)$_2$ cluster (-0.69 eV) (see Table S3 in Supporting Information). Figure 6d depicts interaction of the H atom of H$_2$O molecule with the O atom of the Mg(OH)$_2$(001) surface. The length of the H–O bond of the H$_2$O molecule near to the surface, elongates from 0.972 Å (bare H$_2$O) to 0.994 Å, while another O-H bond is shortened to 0.952 Å. $E_{ads}$ of H$_2$O molecule on the Mg(OH)$_2$(001) surface is found to be -0.37 eV, which is lower than that of the H$_2$O molecule on Mg(OH)$_2$ cluster (-0.74 eV) (see Table S3 in Supporting Information). In addition, the distance between the H$_2$O molecule and the Mg(OH)$_2$(001) surface of 2.017 Å, is longer than that between the H$_2$O molecule and the Mg(OH)$_2$ cluster (1.957 Å). Similar to the case of MgO, the lower $E_{ads}$ of the CO$_2$ and H$_2$O molecules on the Mg(OH)$_2$ cluster, compared to that on the Mg(OH)$_2$(001) surface, suggests stronger interaction of these molecules with the edge and/or defect-containing surface of Mg(OH)$_2$ crystal.

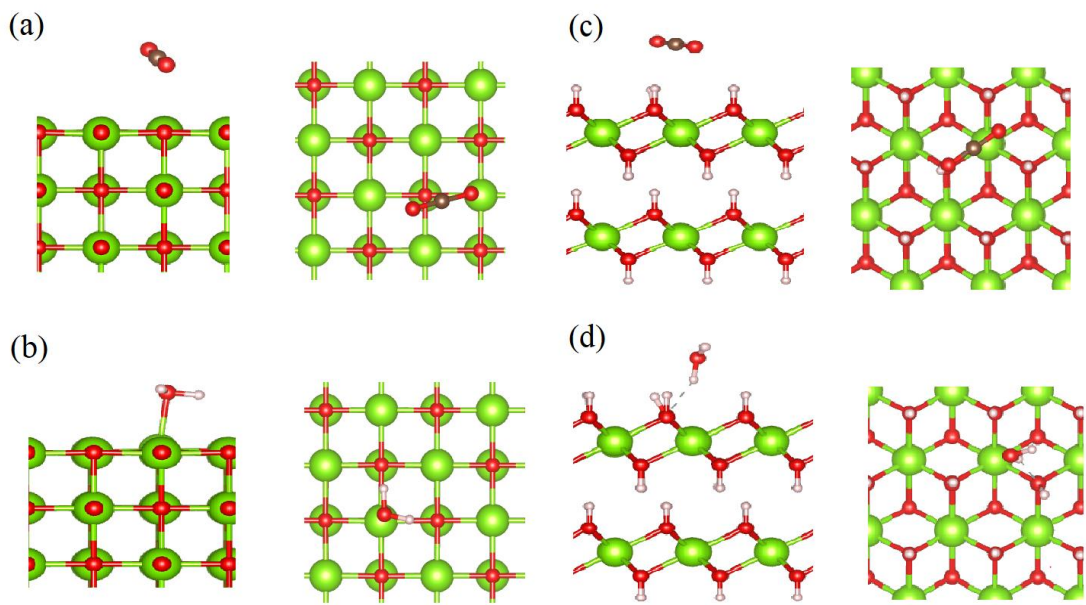

**Figure 6.** The side and top views of the lowest-energy configuration of (a) CO$_2$ and (b) H$_2$O molecules on the MgO(001) surface. The side and top views of the lowest-energy configuration of (c) CO$_2$ and (d) H$_2$O molecules on the Mg(OH)$_2$(001) surface.

**Conclusions**

In the present study, the mechanism of potential reactions on MgO and Mg(OH)$_2$ during HMC synthesis is investigated by DFT-based calculations. The results show that despite the energy barrier for the $CO_2$ molecule adsorption on MgO is lower than that for the $H_2O$ molecule adsorption on MgO, the hydration of MgO is faster, due to the difference in the frequency of $CO_2$ and $H_2O$ molecules collisions with MgO. In addition, it is found that adsorption of $CO_2$ on hydrated MgO is slower than that on bare MgO, which means that the presence of $H_2O$ molecules (moisture environment) can hinder MgO carbonation. In turn, the carbonation of $Mg(OH)_2$ is found to be significantly faster than that of MgO. It should be noted, that both hydration and carbonation of $Mg(OH)_2$ take place at the edges. In addition, a weaker interaction of the $CO_2$ and $H_2O$ molecules with the MgO and $Mg(OH)_2$ surfaces compared to the edge and/or defect-containing surfaces (clusters) is found. Importantly, two limiting factors of the HMCs formation reaction are found: i) surface passivation of $Mg(OH)_2$ upon its initial carbonation and ii) surface covering of $Mg(OH)_2$ by $H_2O$ molecules, which inhibits the carbonation on $Mg(OH)_2$.

**Acknowledgements**
M.G.G., H.N. and P.K. are grateful for the support of the Academy of Finland grant CCC (329477) and the University of Oulu and the Academy of Finland Profi5 funding (326291). K.A.A. and W.C acknowledge funding from the European Research Council (ERC) under the European Union's Horizon 2020 research and innovation programme (grant agreement No. 101002219). H.M. would like to acknowledge funding from "Departamento de Educación, Política Lingüística y Cultura del Gobierno Vasco" (Grant No. IT1358-22). The authors wish to acknowledge CSC–IT Center for Science, Finland, for providing computational resources.